\documentclass[twocolumn,preprintnumbers,nofootinbib,prl]{revtex4}

\usepackage{graphicx}

\usepackage[hypertex]{hyperref}
\newcommand{\beq}{\begin{equation}}
\newcommand{\eeq}{\end{equation}}
\newcommand{\be}{B_\oplus}

\def\be{\begin{equation}}
\def\ee{\end{equation}}
\def\baray{\begin{eqnarray}}
\def\earay{\end{eqnarray}}
\def\ba{\begin{eqnarray}}
\def\ea{\end{eqnarray}}

\begin{document}

\pagestyle{plain}

\preprint{MAD-TH-07-10} \preprint{CERN-PH-TH-2007-166}

\title{Multifield DBI Inflation and Non-Gaussianities}

\author{Min-xin Huang}
\affiliation{Department of Physics, CERN - Theory Division,
CH-1211 Geneva 23, Switzerland }
\author{Gary Shiu}
\affiliation{Department of Physics, University of Wisconsin,
Madison, WI 53706, USA}
\author{Bret Underwood}
\affiliation{Department of Physics, University of Wisconsin,
Madison, WI 53706, USA}


\begin{abstract}

We analyze the trajectories for multifield DBI inflation, which
can arise
 in brane inflation models, and show that the trajectories
are the same as in typical slow roll inflation.
We calculate the
power spectrum and find that the higher derivative terms
of the DBI action
 lead to a suppression of the
contribution from the isocurvature perturbations.  We also
calculate the bispectrum generated by the isocurvature
perturbation, and find that it leads to distinctive features.

\end{abstract}
\maketitle

\section{Introduction}
\setcounter{equation}{0}

Scalar field theories with non-canonical kinetic terms provide 
novel realizations of the inflationary paradigm \cite{kinflation}.
One interesting class of such models which have been studied extensively in recent years is
DBI inflation \cite{DBI,DBISky},
characterized by kinetic terms which arise from the Dirac-Born-Infeld (DBI) action.
A particularly appealing phenomenological feature of DBI inflation is that
 it can lead to strong and unique non-Gaussian signatures 
 in the Cosmic Microwave Background (CMB)
\cite{DBISky,NonGauss}.
The DBI action is ubiquitous in string theory, e.g., as an effective theory for worldvolume degrees of freedom 
on branes.
When embedded in 
brane inflation \cite{BraneInflation},
the inflaton field
in 
DBI models
can be given
a natural geometrical interpretation
as the position of a D-brane in 
extra dimensions.
DBI inflation arises when the D-brane moves in 
a highly warped region of the internal space
where the speed limit is small, and reduces to the usual
slow roll brane inflation (with canonical kinetic term) when the brane is moving 
non-relativisitically with respect to the local warp factor.
Since the position of the brane in each compact direction is described by a scalar field, brane inflation is
naturally a multifield inflationary model.

Multifield models are characterized by their trajectories in field space and can in general be decomposed into
an adiabatic field, which parameterizes motion along the trajectory, and isocurvature fields, which describe
the directions perpendicular to the trajectory.  Features in the trajectory, such as a sharp turn, can convert
isocurvature perturbations into adiabatic/curvature perturbations (even on superhorizon scales) and can give
rise to interesting features in the primordial power spectrum and non-Gaussianity.

In this paper we study the effects of multiple fields in DBI inflation.  In particular, we study the multifield DBI trajectories
and show that they are identical to the usual slow-roll multifield case in which the trajectory is dominated by
the field with the largest slope of the potential.  We calculate the power spectrum for multiple DBI fields in the limit the
trajectory makes a sharp turn and show that the contribution of the isocurvature perturbations to the power spectrum is suppressed
by the sound speed.  Finally, we calculate the non-Gaussianity in the sharp turn limit and find that the non-Gaussianity
is dominated at leading order in the sound speed by the usual single field DBI contribution, but has new multifield features
at sub-leading order.  We conclude by commenting on the impact of our results for DBI model building.

\section{Multifield Equations of Motion}

Consider a 10-dimensional warped throat background with the metric
\begin{equation}
ds^2 = \tilde{f}^{-1/2}(y) g_{\mu\nu}dx^{\mu}dx^{\nu} + \tilde{f}^{1/2}(y)\tilde{g}_{mn}dy^m dy^n
\end{equation}
common to type IIB string compactifications \cite{GKP,KS},
where $\tilde{f}(y)$ is called the warp factor of the throat which can in principle depend on all of the
coordinates of the internal space $\vec{y}$.

A D3-brane in this background is described by (to lowest order in string coupling and to all orders in
$\alpha'$),
\begin{eqnarray}
&&S_{DBI} = -\int d^4 x \sqrt{-g}\, [ \frac{1}{f(\phi_i)} \nonumber \\
&&  \left(\sqrt{1+f(\phi_i)g^{\mu\nu}\sum_i\partial_{\mu}\phi_i\partial_{\nu}\phi_i}-1\right)
            + V(\phi_i)]
\label{eq:DBIAction}
\end{eqnarray}
where the warp factor is rescaled $f(\phi_i) = T_{D3}^{-1} \tilde{f}(y(\phi_i))$, and the
real canonical scalar fields associated with the motion of the brane are given by
\begin{equation}
\phi_i \equiv T_{D3}^{1/2} y_i\,
\end{equation}
where $i=1,...6$.
The potential $V(\phi_i)$ can arise, for example, from interactions with 
$\overline{D}3$-branes, $D7$-branes, or from
the breaking of the local isometries of the compact space, and we will leave
it to be unspecified for the moment.

We will define the sound speed during inflation to be the
inverse of the
``Lorentz factor" of the DBI action for spatially homogeneous
fields,
\begin{eqnarray}
c_s=\sqrt{1-f(\phi_i)\sum_i \dot{\phi}_i^2}
\end{eqnarray}
We will be interested in the small sound speed limit $c_s\ll 1$
where the non-Gaussianity is observable. In the following we will
use the convention that the $y_i$ measure the distance from IR ``tip"
of the throat and that the D3 brane moves towards the tip, so $\dot{\phi}_i<0$.

Consider a FRW universe with
four-dimensional metric,
\begin{eqnarray}
ds^2=-dt^2+a(t)^2\sum_{i=1}^3dx_i^2.
\end{eqnarray}
The Friedman equation and equations of motion for the fields
$\phi_i$ are (with $H = \frac{\dot{a}}{a}$)
\begin{eqnarray}
&&3M_p^2H^2=\frac{1}{f(\phi_i)}\left(\frac{1}{c_s}-1\right)+V(\phi_i), \\
&& \frac{1}{a(t)^3}\frac{d}{dt}(a(t)^{3}\frac{\dot{\phi}_i}{c_s}) = \nonumber \\
    &&\ \ -\frac{\partial}{\partial\phi_i}\left(V(\phi_i)+ f(\phi_i)^{-1}(c_s-1)\right)
\label{eq:EOM}
\end{eqnarray}
Distributing the time derivative, the equation of motion
(\ref{eq:EOM}) can also be written
\begin{equation}
\ddot{\phi}_i + 3 H \dot{\phi}_i -
\frac{\dot{c}_s}{c_s}\dot{\phi}_i+c_s \partial_{\phi_i}\left(V+\frac{(c_s-1)}{f} \right) =0\, .
\label{eq:EOM2}
\end{equation}
Clearly the equation of motion for a homogeneous scalar field with
a canonical kinetic term is obtained from
(\ref{eq:EOM}),(\ref{eq:EOM2}) in the limit $c_s\rightarrow 1$.

We define the {\em multifield DBI inflationary parameters} as \cite{NonGauss}
\begin{eqnarray}
\label{eq:MultifieldDBI2}
\epsilon &\equiv & -\frac{\dot{H}}{H^2} \\
\eta_{ij} &\equiv& M_p^2 c_s \frac{\partial_{\phi_i}\partial_{\phi_j}V}{V}\, .
\label{eq:MultifieldDBI}
\end{eqnarray}
When these parameters are much smaller than one, the equations of motion (\ref{eq:EOM2}) take a form similar to that
of a slowly rolling field
\begin{eqnarray}
&&3H\dot{\phi_i}\left(1-\frac{\sum_j\tan\theta_{ij}\eta_{ij}+\epsilon}{3}\right) \nonumber \\
    &&+\partial_{\phi_i}\left(V(\phi_i)+f^{-1} (c_s-1)\right)\nonumber \\
&&\approx
 3 H \dot{\phi}_i + c_s \partial_{\phi_i}\left(V(\phi_i)+f^{-1} (c_s-1)\right) = 0\, ,
\label{eq:EOMLimit}
\end{eqnarray}
where $\tan\theta_{ij} \approx \partial_{\phi_j}V/\partial_{\phi_i}V$ to leading order in the DBI inflationary parameters 
(\ref{eq:MultifieldDBI2}-\ref{eq:MultifieldDBI}).

As a specific example, for the ``standard case" of an AdS warp factor that depends only on one of the fields
$f(\phi_1) = \lambda/\phi^4$ and a separable potential of mass terms $V = \frac{1}{2}\sum m_i^2\phi_i^2$,
it can be shown that for small sound speed $c_s\ll 1$ and a large mass hierarchy ($m_i\gg m_j$ for some $i,j$)
the multifield DBI inflationary parameters are
of order
${\mathcal O}(\epsilon)$.

\section{Trajectories in Multifield Brane Inflation}

To simplify our analysis we will restrict ourselves to a two field model $(\phi_1,\phi_2)$,
but
it is straightforward
to generalize
our analysis
 to any number of fields.

\begin{figure}[t]
\begin{center}
\includegraphics[scale=.5]{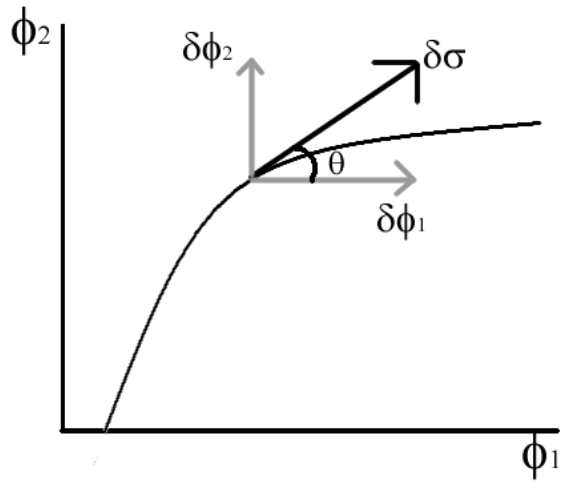}
\caption{\small The trajectory of a multiple field inflationary system can be decomposed into an ``adiabatic" field $\sigma$ with components
along the trajectory and an ``entropy" field $s$ orthogonal to the trajectory.}
\label{fig:Adiabatic}
\end{center}
\end{figure}

We will parameterize the classical trajectory by an ``adiabatic field" $\sigma$ that represents the component of the
field motion along the trajectory \cite{Multifield} (see Figure \ref{fig:Adiabatic}),
\begin{equation}
\dot{\sigma} = (\cos\theta)\dot{\phi}_1+(\sin\theta)\dot{\phi}_2\, .
\end{equation}
The angle $\theta$ parameterizes the angle the classical trajectory makes with one of the field directions (here
chosen to be $\phi_1$) and should
not be confused with the angular position of the D3 brane in the compact space, which is parameterized by the fields $\phi_i$.
The ``entropy field" $s$ transverse to the classical trajectory gives
rise to isocurvature fluctuations which are given by,
\begin{equation}
\delta s = -(\sin\theta)\delta \phi_1+(\cos\theta)\delta\phi_2\, .
\end{equation}
By definition, the entropy field is constant, $\dot{s} = 0$.

Using this parameterization, we can rewrite the exact equations of motion (\ref{eq:EOM2}) as (in particular,
the multifield DBI parameters are not necessarily assumed small for this expression),
\begin{eqnarray}
\frac{\ddot{\sigma}}{c_s}&+&\frac{3 H \dot{\sigma}}{c_s}-\frac{\dot{c}_s}{c_s^2}\dot{\sigma}
+\partial_{\sigma}(V+f^{-1}(c_s-1)) = 0\ \ \\
\dot{\theta} \dot{\sigma}&=& \partial_{\phi_1}(V+f^{-1}(c_s-1))\sin\theta -\nonumber \\
&&\partial_{\phi_2}(V+f^{-1}(c_s-1))\cos\theta\, .
\label{eq:AngleEOM}
\end{eqnarray}
It is clear from (\ref{eq:AngleEOM}) that the angle of the trajectory has a ``fixed point" trajectory in field space
$\dot{\theta}\approx 0$ defined by
\begin{equation}
\tan\theta_* = \frac{\partial_{\phi_2}(V+f^{-1}(c_s-1))}{\partial_{\phi_1}(V+f^{-1}(c_s-1))}\, .
\label{eq:Angle}
\end{equation}
Furthermore this fixed point is stable to leading order when the slope of the potential is positive and 
dominates over the slope of the warp factor 
since small variations
$\delta \theta$ are driven to zero (notice that $\dot{\sigma} < 0$).
This fixed point can also be seen from the 
equations of motion (\ref{eq:EOMLimit}) in the ``DBI slow roll regime" 
(e.g. when the multifield DBI parameters (\ref{eq:MultifieldDBI2}-\ref{eq:MultifieldDBI}) are small),
\begin{equation}
\tan \theta = \frac{\dot{\phi}_2}{\dot{\phi}_1} \approx \frac{\partial_{\phi_2}(V+f^{-1}(c_s-1))}{\partial_{\phi_1}(V+f^{-1}(c_s-1))}\, .
\label{eq:ThetaLimit}
\end{equation}
We see then that being in the DBI slow roll regime is equivalent to being at the stable ``fixed point" of the trajectory, thus
DBI slow roll is an attractor solution.  

Let us examine (\ref{eq:Angle}) in more detail.  First, we note that for an inflationary solution we require
that the potential energy dominates over the kinetic energy, e.g. $V(\phi_i)\gg 1/(c_s f)$, so that for small $c_s$, $V(\phi_i)\gg c_s/f$ is
automatically true.  The trajectories (\ref{eq:Angle}) are now the same as in standard multifield inflation, in particular,
the angle of the trajectory is controlled by the ratio of the curvatures of the potential: the trajectory follows the direction with the largest
curvature.  When the field with the largest curvature reaches its minimum the trajectory makes a sharp turn in field space, with the sharpness
of the turn given by the ratio of the curvatures.

\begin{figure}[h]
\begin{center}
\includegraphics[scale=.4]{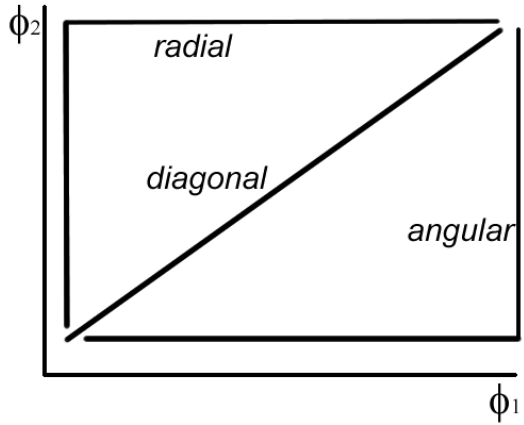}
\caption{\small Multifield models have a number of different trajectories, depending on the curvature of the
potential for the fields.  The sharpness of the turns is controlled by the ratio of the curvatures.}
\label{fig:Trajectory}
\end{center}
\end{figure}

If we express the warped geometry as a cone over an angular base space $X^5$,
\begin{equation}
\tilde{g}_{mn}dy^m dy^n = dr^2+ds^2_{X^5}
\end{equation}
we can identify one of the fields $\phi_1$ as the radial
coordinate and the other field $\phi_2$ as one of the angular
coordinates on the base $X^5$ of the D-brane.  In this case we can
roughly classify the trajectories based on which field dominates
at early times, as shown in Figure \ref{fig:Trajectory}: a {\it radially dominated} trajectory is when the
slope of the potential in the angular direction $\phi_2$ is much smaller than
the radial direction $\phi_1$ and so the trajectory is dominated
at early times by motion in the radial direction; a {\it diagonal}
trajectory is when the slopes of potentials of the two fields are
approximately the same and so the trajectory is approximately a
diagonal line composed of a linear combination of the radial and
angular directions; and a {\it angularly dominated} trajectory is
when the slope of the potential in the radial direction $\phi_1$ is the
smallest so the trajectory is dominated at early times by motion
in the angular $\phi_2$ direction. Clearly the diagonal trajectory
does not have a significant turn in field space, and so will not
contribute to a generation of curvature perturbations as discussed
above.  In fact, the diagonal-type trajectory is just a
linear combination of the fields and hence can be completely
described by a single field, and so we will not consider this
possibility.

\section{The Power Spectrum of Multifield DBI Inflation}

The equation of motion for the curvature perturbation ${\mathcal R}\approx \zeta$ in a multifield model with fields $\phi_I$ can be solved
exactly using the $\delta N$ formalism
\cite{DeltaN,2FNonGauss}, which states that the curvature perturbation is equal to the difference between the number of e-folds of the classical
trajectory $N_e$ and the perturbation to the classical trajectory ${\mathcal N}_e$,
\begin{equation}
\zeta = dN_e = {\mathcal N}_e - N_e = \sum_i N_{,i} (\delta \phi_i)^*\, ,
\label{eq:deltaN}
\end{equation}
where $N_{,i}\equiv \frac{\partial N_e}{\partial \phi_i^*}$ is the
derivative of the number of e-foldings with respect to the field
evaluated when the mode exits the horizon $\phi_i^*$. For two
fields, the power spectrum is then given by,
\begin{eqnarray}
4\pi^2P_{\zeta} &=& \langle \zeta \zeta\rangle = (N_{,1})^2 \langle
\phi_1\phi_1\rangle \nonumber \\
&&+ 2 (N_{,1})(N_{,2})\langle
\phi_1\phi_2\rangle + (N_{,2})^2 \langle \phi_2\phi_2\rangle\, ,
\label{eq:GeneralPowerSpectrum}
\end{eqnarray}
where we have allowed for cross correlation between the fields - for a canonical kinetic term, the cross coupling is zero.  We see, then, that the
power spectrum receives extra contributions in multifield inflation, 
both from cross couplings and from the extra two point correlation functions of the additional fields.

We would like to analyze these extra contributions in more detail.
Using the approximation that the DBI multifield parameters (\ref{eq:MultifieldDBI}) are small and that the potential is
separable $V(\phi_1,\phi_2) = V_1(\phi_1)+V_2(\phi_2)$ and dominates the energy
density, the number of e-folds from the time of horizon crossing to the end of inflation is \cite{2FNonGauss}
\begin{eqnarray}
N_e &=& -\frac{1}{M_p^2}\int_*^e \frac{V_1}{c_s
    \partial_{\phi_1}V_1} d\phi_1  \nonumber \\
&-&\frac{1}{M_p^2}\int_*^e
    \frac{V_2}{c_s \partial_{\phi_2}V_2} d\phi_2\, . \label{eq:Efolds}
\end{eqnarray}
As in \cite{2FNonGauss}, we can write the curvature perturbation (\ref{eq:deltaN}) in a much simpler way by using a different
set of multifield DBI parameters,
\begin{equation}
\epsilon_i = \frac{c_s M_p^2}{2}\left(\frac{V_i'}{V}\right)^2
\end{equation}
where a prime denotes a derivative with respect to the
argument. Note that $\epsilon = -\dot{H}/H^2 =
\epsilon_1+\epsilon_2$.  The curvature perturbation can now be
written,
\begin{eqnarray}
\zeta = &&\frac{1}{M_p \sqrt{2 \epsilon_1^*}}\left(\frac{V_1^*+Z^e}{V^*}\right)(\delta \phi_1)^*\nonumber \\
            &&+ \frac{1}{M_p \sqrt{2 \epsilon_2^*}}\left(\frac{V_2^*-Z^e}{V^*}\right)(\delta\phi_2)^*\, ,
\label{eq:2FCurvature}
\end{eqnarray}
with
\begin{equation}
Z^e \equiv \frac{V_2^e \epsilon_1^e-V_1^e \epsilon_2^e}{\epsilon^e}
\end{equation}
where the superscript `e' denotes evaluation at the end of inflation.
In the limit that the trajectory is in the {\em radial} or {\em angular} direction as shown above then at the end of inflation
the one of the field has settled into its minima (say, $\phi_2$ for concreteness) so $Z^e = V_2^e = \mbox{const}$.  For mass-term
dominated potentials (or equivalently when the vacuum energy from $\phi_2$ is small $V_2^e \ll V_2^*$) then we can take
$Z^e = 0$ and the expression for the curvature perturbation simplifies to include
only the values of the potential and the slow roll parameters evaluated at horizon crossing,
\begin{eqnarray}
\zeta &=& \frac{1}{M_p \sqrt{2 \epsilon_1^*}}\left(\frac{V_1^*}{V^*}\right)(\delta \phi_1)^*\nonumber \\
            &+& \frac{1}{M_p \sqrt{2 \epsilon_2^*}}\left(\frac{V_2^*}{V^*}\right)(\delta\phi_2)^*\, .
\label{eq:CurvatureSimple}
\end{eqnarray}
This simple expression for the curvature perturbation will be useful later in evaluating the power spectrum.

\subsection{The inflationary perturbation}

In order to calculate the power spectrum we need to evaluate the two point correlation functions for the
perturbations of the scalar fields in (\ref{eq:GeneralPowerSpectrum}).
In order to have analytic control over our expressions we will assume that the trajectory is highly {\em radial}, e.g.
\begin{equation}
\tan\theta = \frac{\dot{\phi}_2}{\dot{\phi}_1} = \frac{V_2'}{V_1'} \ll c_s\, .
\end{equation}
Under this assumption, the adiabatic perturbation is simply the perturbation in the $\phi_1$ direction
and perturbations in the $\phi_2$ direction are just isocurvature perturbations,
\begin{equation}
\delta \sigma=\delta\phi_1,~~~~ \delta s= \delta\phi_2\, .
\end{equation}

Expanding the kinetic part of the Lagrangian in (\ref{eq:DBIAction}) to
quadratic order in $\delta \sigma$ and $\delta s$, we find
\begin{eqnarray}
\mathcal{L}_2  &=& \frac{a^3}{2c_s^3}[\dot{\delta\sigma
    }^2-a^{-2}c_s^2(\nabla\delta\sigma)^2] \nonumber \\
&+&\frac{a^3}{2c_s}[\dot{\delta s}^2-a^{-2}(\nabla \delta s)^2]
\label{eq:QuadLagrangian}
\end{eqnarray}
Note that the isocurvature fluctuations (e.g. the fluctuations in the angular
direction) scale differently with the sound speed; we will see soon that this
has important consequences for the two point functions.

The quantization of the perturbations proceeds as usual with
\begin{eqnarray}
\delta\sigma(\tau,\textbf{x})&=&\frac{1}{(2\pi)^3}\int d^3k\nonumber \\
&&\left[u(\tau,{\textbf{k}})a(\textbf{k}) +u^*(\tau,{-\textbf{k}})a^{\dagger}(-\textbf{k})\right]e^{i\textbf{k}\cdot\textbf{x}}~, \nonumber \\
\textrm{where}~~~ &&
u(\tau,{\textbf{k}})=\frac{H}{\sqrt{2k^3}}(1+ikc_s\tau)e^{-ikc_s\tau}
\end{eqnarray}
for the adiabatic perturbation and
\begin{eqnarray}
\delta s(\tau,\textbf{x})&=&\frac{1}{(2\pi)^3}\int d^3k \nonumber \\
&&\left[v(\tau,{\textbf{k}})b(\textbf{k})+v^*(\tau,{-\textbf{k}})b^{\dagger}(-\textbf{k})\right]e^{i\textbf{k}\cdot\textbf{x}}~,\nonumber \\
\textrm{where}~~~ &&
v(\tau,{\textbf{k}})=H\sqrt{\frac{c_s}{2k^3}}(1+ik\tau)e^{-ik\tau}
\end{eqnarray}
for the isocurvature modes.
The creation and annihilation operators satisfy the usual
commutation relation
$[a(\textbf{k}),a^{\dagger}(\textbf{k}^{\prime})]=[b(\textbf{k}),b^{\dagger}(\textbf{k}^{\prime})]=
(2\pi)^3\delta^3(\textbf{k}-\textbf{k}^{\prime})$,
$u(\tau,{\textbf{k}})$ and $v(\tau,{\textbf{k}})$ are the
solutions of the quadratic Lagrangian whose normalizations are
fixed by the Wronskian conditions \footnote{This can be checked by
computing the commutator
$[\delta\sigma(\tau,\textbf{x}_1),p_{\delta{\sigma}}(\tau,\textbf{x}_2)]=i\delta^3(\textbf{x}_1-\textbf{x}_2)$,
where $p_{\delta{\sigma}}=\frac{\partial{{\mathcal L}_2}}{\partial
\dot{\delta\sigma}}$ is the canonical momentum.}, and
$\tau=-\frac{1}{aH}$ is the conformal time.

It is now straightforward to calculate the two point functions,
\begin{eqnarray}
\label{eq:2ptadiabatic}
\langle\delta\sigma (\textbf{k}_1)\delta\sigma( \textbf{k}_2)\rangle&=&
(2\pi)^3\delta^3(\textbf{k}_1+\textbf{k}_2)\frac{H^2}{2k_1^3},
\\
\langle \delta s( \textbf{k}_1) \delta s( \textbf{k}_2)\rangle &=&
(2\pi)^3\delta^3(\textbf{k}_1+\textbf{k}_2)\frac{c_sH^2}{2k_1^3}.
\label{eq:2ptiso}
\end{eqnarray}
Here
$\delta\sigma(\textbf{k})=\delta\sigma(\tau,\textbf{k})|_{\tau\rightarrow
0}$, $\delta s(\textbf{k})=\delta
s(\tau,\textbf{k})|_{\tau\rightarrow 0}$, and the Hubble parameter
and sound speed are evaluated at the time of horizon crossing.
Here we see  that the isocurvature fluctuations generated by
$\delta s$ are suppressed by a factor of $c_s \ll 1$ compared to
the adiabatic perturbations generated by $\delta \sigma$. This
suppression can be traced back to the different $c_s$ dependence
found in the quadratic Lagrangian (\ref{eq:QuadLagrangian}).
We note that the exact same calculation follows through if the
trajectory is dominated by the {\em angular} direction after the
replacement $\delta\phi_1\leftrightarrow \delta\phi_2$.
Finally, notice also that in the limit of a straight-line trajectory there is no
cross correlation between the fields, so the cross term in the
power spectrum (\ref{eq:GeneralPowerSpectrum}) vanishes.  
Unfortunately, non-straight line trajectories are beyond our analytic control
so it is not clear if the cross coupling 
will be significant, although we expect
that in the diagonal limit the two point function should reduce to that of a simple single
field model where the results are well known.

Using the two point functions
(\ref{eq:2ptadiabatic},\ref{eq:2ptiso}) and the expression for the
curvature perturbation (\ref{eq:CurvatureSimple}), the power
spectrum (\ref{eq:GeneralPowerSpectrum}) becomes,
\begin{eqnarray}
P_{\zeta} &=& \frac{H^2}{4\pi^2
M_p^2}\left[\frac{1}{2\epsilon_1^*}\left(\frac{V_1^*}{V^*}\right)^2+\frac{c_s}{2\epsilon_2^*
}\left(\frac{V_2^*}{V^*}\right)^2\right]
 \nonumber \\
    &\approx & \frac{H^2}{4\pi^2 M_p^2}\frac{1}{2\epsilon_1^*}\left(\frac{V_1^*}{V^*}\right)^2
\label{eq:PowerSpect}
\end{eqnarray}
where in the last line we assume that
the second term is small compared to the first term in the small
$c_s$ limit.

Since the contribution of the angular modes to the power spectrum is highly suppressed
by the sound speed we find that multifield DBI reduces essentially
to the single field case, in contrast to multiple
field slow roll inflation where additional fields may become important when the trajectory makes sharp
turns in field space \cite{2FNonGauss}.

\subsection{Multifield Non-Gaussianity}

In the following we study the non-Gaussianities in more details,
and as it turns out there are some potentially observable
differences from the single field DBI inflation.  To compute the
non-Gaussianities, we expand the DBI Lagrangian to higher order.
The leading order and subleading order cubic terms are
\begin{eqnarray}
\label{cubicterm}
\mathcal{L}_3 &=&
\frac{a^3}{2c_s^5\dot{\sigma}}[\dot{\delta\sigma}^3-a^{-2}c_s^2\dot{\delta\sigma}(\nabla\delta\sigma)^2\nonumber \\
&-&a^{-2}c_s^2\dot{\delta\sigma}(\nabla \delta s )^2]
\end{eqnarray}
The leading contribution to the non-Gaussianity comes from the
first two terms, and their size  is well known
\begin{eqnarray}
f_{NL}\sim \frac{1}{c_s^2}
\end{eqnarray}

Since the angular mode $\delta s$ is suppressed by a factor of
$\sqrt{c_s}$ comparing to the radial mode $\delta \sigma$, we see
that the third term in (\ref{cubicterm}) contributes a
non-Gaussianity of order
\begin{eqnarray}
f_{NL}\sim \frac{1}{c_s}
\end{eqnarray}
Although it is sub-leading comparing to the first two terms, it is
still
potentially observable by future experiments if the sound speed
$c_s$ is small enough. In particular, this effect can be larger
than the sub-leading effect of order $\frac{\epsilon}{c_s^2}$
computed in \cite{NonGauss} when the sound speed satisfies $c_s>
\epsilon$.

The calculation of the
three-point function is standard, see e.g. \cite{NonGauss} for
details,
\begin{eqnarray}
&&\langle \zeta(\textbf{k}_1)\zeta(\textbf{k}_2)
    \zeta(\textbf{k}_3)\rangle = -i(2\pi)^3\delta^3(\sum_i\textbf{k}_i) \nonumber \\
&&\times v(0,\textbf{k}_1)v(0,\textbf{k}_2)u(0,\textbf{k}_3) \left(\frac{\partial
    N}{\partial \phi_1^*}\right)\left(\frac{\partial N}{\partial \phi_2^*}\right)^2  \nonumber \\
&&\times\frac{1}{c_s^3\dot{\sigma}}\int _{-\infty}^0a d\tau \frac{d u^*(\tau, \textbf{k}_3)}{d\tau}
    [(-\textbf{k}_1\cdot\textbf{k}_2)v^*(\tau,\textbf{k}_1)v^*(\tau,\textbf{k}_2)] \nonumber \\
&&+\textrm{c.c.}+\textrm{cyclic}.
\end{eqnarray}
here ``cyclic.'' means two other terms by cyclically permuting
$k_1, k_2, k_3$. We can see the non-Gaussianity vanishes in the
squeezed limit when one of the momentum $k_i\rightarrow 0$, since
$\frac{d u^*(\tau, \textbf{k}_3)}{d\tau} \sim k_3^2$ and there is
factor of $\textbf{k}_1\cdot\textbf{k}_2$. This is the same as in
single field DBI inflation.

\begin{figure}
\begin{center}
\includegraphics[scale=.5]{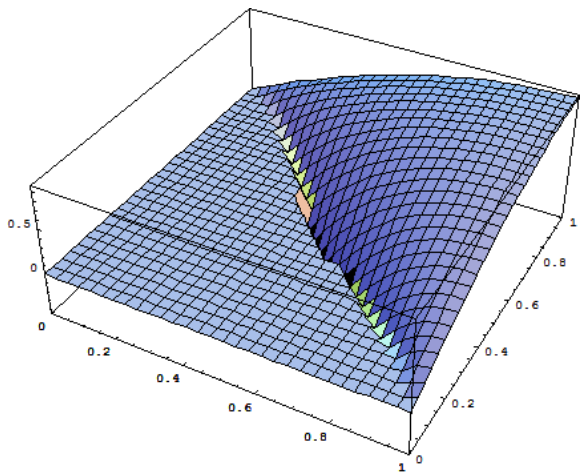}
\caption{The shape of non-Gaussianity in multifield DBI inflation is shown through a plot
of ${\mathcal A}(k_1,k_2,1)/(k_1k_2)$ in (\ref{eq:NGShape}).  Notice that in the folded triangle limit $k_1=k_2=k_3/2$
the bispectrum is negative, and constitutes a distinctive signature of multifield DBI inflation in the small $c_s$ limit.  The
presence of opposite signs of the non-Gaussianity may give rise to interesting observational effects.}
\label{fig:nongauss}
\end{center}
\end{figure}

Going away from the squeezed limit, we can compute the above three
point function assuming $k_1$, $k_2$, $k_3$ are of the same order
of magnitude. The non-Gaussianity has a very interesting shape as
the following
\begin{eqnarray}
\label{eq:NGShape}
\mathcal{A}(k_1,k_2,k_3)&=&\frac{k_3^2(k_1^2+3k_1k_2+k_2^2)(-\textbf{k}_1\cdot\textbf{k}_2)}{(k_1+k_2)^3}\ \ \\
&&+\textrm{cyclic}. \nonumber
\end{eqnarray}
We plot the non-Gaussianity as $\mathcal{A}(k_1,k_2,1)/(k_1k_2)$
following the convention of \cite{NonGauss} in Figure
\ref{fig:nongauss}.

\begin{figure}
\begin{center}
\includegraphics[scale=.6]{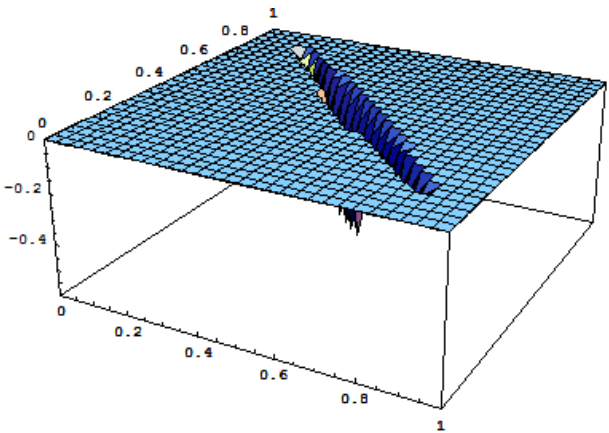}
\caption{The negative part of the non-gaussianity in the folded triangle limit is shown.}
\label{fig:NonGaussNeg}
\end{center}
\end{figure}

We can see from Figure \ref{fig:NonGaussNeg} that the non-Gaussianity has a very
interesting new feature, namely, the sign of non-Gaussianity is
different in the middle of the ``folded triangle'' limit where
$k_3=k_1+k_2$ from most of the other region of the configuration
space. For example, one can directly check for the configuration
of a folded triangle $k_1=k_2=\frac{k_3}{2}$, the shape is
negative $\mathcal{A}(k_1,k_2,k_3)=-\frac{1.088}{k_3^3}$. This
feature is not present in any other known inflationary models, so
it can be used as a distinctive signature of multifield DBI
inflation. 
The sign of non-Gaussianity determines the sense of skewness in the CMB temperature and matter density, and thus this change in sign can lead to potentially interesting observational effects \footnote{The convention for the sign of $f_{NL}$ (which characterizes the level of non-Gaussianity) has been a source of confusion in the literature. See \cite{Komatsu} for a discussion.}.
However, we should caution that since this is a
sub-leading effect, it might be difficult to disentangle the
signature from experimental data.
It would be interesting to extend this analysis to more than two fields,
which may lead to additional enhancements or suppressions of the multifield 
non-Gaussianities \cite{Nflation}.

\subsection{Discussion}

We have shown that
the two point function of the extra ``angular" scalar field direction
during inflation is suppressed by a factor of $c_s$ compared to the usual single field contribution,
thus
the multifield DBI observables simply reduce to the single field case (in the limit where
one of the fields dominates the trajectory).
There have been a lot of recent
attempts in
 building single field DBI inflation models
consistent with all known compactification constraints and precision
cosmological observations \cite{DBIModelBuilding}.
In particular, these works find that combining observational constraints of the
amplitudes and tilt of the scalar and tensor perturbations and the primordial
non-gaussianity together with limitations on the field range coming from compactification
puts severe constraints on the viable parameter space of single field DBI models, ruling
out the most simple models.

In general, since we see that the small sound speed in multifield DBI inflation suppresses
the multifield effects,
multifield DBI inflationary observables can be well approximated by
their single field values.
Note that this also implies that
the dramatic ${\mathcal O}(1)$ effects expected at the end of multifield DBI inflation
due to the inhomogenous surface of tachyon condensation
examined in \cite{Leblond} are now instead suppressed by ${\mathcal O}(c_s)$ and
are subdominant.

Variations of the basic DBI model, such as its IR version  \cite{Chen, Thomas}, 
models involving wrapped branes \cite{WrappedDBI} or different warped geometries, 
may be able to evade the strong constraints of \cite{DBIModelBuilding}; since we
have not made explicit use of any particular model we expect our results to hold
in the small sound speed limit of these models as well.

{\it Note: As this manuscript was being prepared, a preprint \cite{Spinflation} 
appeared which contains some overlap with this work.}

\vspace{0.2in} {\leftline {\bf Acknowledgments:}}

\medskip

We would like to thank Thorsten Battefeld, Damien Easson, Louis Leblond, Liam McAllister, Sarah
Shandera and Gianmassimo Tasinato for helpful discussions and comments. The work of MH, GS and
BU was supported in part by NSF CAREER Award No. PHY-0348093, DOE
grant DE-FG-02-95ER40896, a Research Innovation Award and a
Cottrell Scholar Award from Research Corporation.

\end{document}